\documentclass[12pt,a4]{ronbun}

\usepackage{color,amsmath,amssymb,graphicx}

\usepackage{cite}
\usepackage{bm}
\usepackage{dcolumn}

\newcommand{\s}{\scriptscriptstyle}

\newcommand{\pd}{\partial}

\newcommand{\nn}{\nonumber}
\newcommand{\e}{{\rm e}}

\newcommand{\del}{\delta}

\newcommand{\al}{\alpha}

\renewcommand{\th}{\theta}

\newcommand{\ba}{\begin{eqnarray*}}
\newcommand{\ea}{\end{eqnarray*}}

\newcommand{\KUCPlogo}{\hbox{\lower 1.4ex\hbox{
\Huge\boldmath $\cal K$}
\kern -1.15em {\sffamily \bfseries\large\ UCP}}
\kern -4.5em \raise 0.2em\hbox{\lower 1.4ex\hbox{\color{cyan}
\Huge\boldmath $\cal K$}
\kern -1.15em {\color{magenta}\sffamily \bfseries\large\ UCP}
\put(-20,-7){\tiny\it preprint}
}}

\numberwithin{equation}{section}

\begin{document}

\begin{flushright}

\parbox{3.2cm}{
{KUCP-0211 \hfill \\
{\tt hep-th/0206070}}\\
\date
 }
\end{flushright}

\vspace*{0.5cm}

\begin{center}
 \Large\bf Supermembrane on the PP-wave Background

\end{center}

\vspace*{1.0cm}

\centerline{\large Katsuyuki Sugiyama$^{\ast}$ 
and Kentaroh Yoshida$^{\dagger}$}

\begin{center}
$^{\ast}$\emph{Department of Fundamental Sciences, \\
Faculty of Integrated Human Studies, \\
Kyoto University, Kyoto, 606-8501, Japan.} \\
{\tt E-mail:sugiyama@phys.h.kyoto-u.ac.jp}\\
\vspace{0.2cm}
$^{\dagger}$\emph{Graduate School of Human and Environmental Studies,
\\ Kyoto University, Kyoto 606-8501, Japan.} \\
{\tt E-mail:~yoshida@phys.h.kyoto-u.ac.jp}
\end{center}

\vspace*{1.2cm}

\centerline{\bf Abstract}

We study the closed and open supermembranes on the maximally supersymmetric 
pp-wave background. 
In the framework of
the membrane theory, the superalgebra is calculated by
using the Dirac bracket and we obtain its central extension by 
surface terms. The result supports the existence of the extended objects
in the membrane theory in the pp-wave limit.
When the central terms are discarded, 
the associated algebra completely agrees with that of
Berenstein-Maldacena-Nastase matrix model. 
We also discuss the 
open supermembranes on the pp-wave and elaborate 
the possible boundary conditions.

\vspace*{1.5cm}

\thispagestyle{empty}
\setcounter{page}{0}

\newpage 

\section{Introduction}

For the past years, many works toward the 
investigation of the M-theory 
has been done, and in particular the matrix model approach 
seems greatly successful \cite{BFSS}. 
These are attempts to describe the scattering in the
eleven dimensional supergravity theory.
For many years, the eleven-dimensional 
supergravity backgrounds have been studied, and the Minkowski-space, 
$AdS_4 \times S^7$, $AdS_7\times S^4$ and Kowalski-Glikman (KG) 
pp-wave solution \cite{KG} are only known cases as the maximally
supersymmetric backgrounds. They are possible candidates 
for the M-theory backgrounds. In particular, by taking a certain limit 
called Penrose limit 
\cite{Penrose,Guven}, the 
KG solution can be obtained from the $AdS_{4}\times S^7$ 
or $AdS_7 \times S^4$ backgrounds \cite{OP}.  
Also, the maximally supersymmetric 
IIB supergravity background has been lately found \cite{OP2}
and it has been shown that the Green-Schwarz (GS) 
type IIB superstring theory on the pp-waves is exactly 
solvable \cite{M,MT,RT}. The pp-wave background used 
in the works \cite{M,MT,RT} can be also obtained by taking Penrose limit 
in the $AdS_5\times S^5$ \cite{OP2}. 

The fact that the maximally supersymmetric pp-wave backgrounds are obtained 
by taking the Penrose limit in the $AdS$ background leads to the work 
\cite{Malda}, where the IIB string theory on the pp-wave 
is used for investigating the $AdS$/CFT correspondence
\cite{Mald,ooguri} 
in the string theoretic analysis. 
That is the exactly solvable model with nontrivial 
string background and it provides an interesting 
area to study properties of strings with 
background fluxes.
Moreover, 
Banks-Fischler-Shenker-Susskind (BFSS) matrix model 
on the maximally supersymmetric pp-wave 
(which we refer to Berenstein-Maldacena-Nastase (BMN) matrix
model) has been proposed from the considerations for the
superparticles. The action of the BMN matrix model has been also 
derived directly from the membrane theory on the maximally supersymmetric 
pp-wave \cite{DSR}.

In this paper we consider the closed and open supermembranes 
on the eleven-dimensional maximally supersymmetric 
pp-wave background. We calculate the supercharges and associated
algebra. In contrast with the algebra in the BMN model, 
surface terms are included in our membrane case. 
It is the central extension of the superalgebra in the BMN model
and we discuss the 
extended objects contained in the membrane theory on the pp-wave. 

Next we discuss the boundary conditions for the 
open supermembrane on the pp-wave by calculating surface terms 
under the variations of 
the supersymmetry transformations. In the case of flat background, 
the open supermembrane
can end on the $p$-dimensional hypersurface only for the values 
$p=1,\,5$ and 9. However, we show that 
some additional surface terms arise in the pp-wave case
and only the value $p=1$ is
allowed for the open supermembrane on the pp-wave. 

This paper is organized as follows. In section 2, as a short review we 
provide an explanation of 
the action of the supermembrane and supersymmetries on
the maximally supersymmetic pp-wave background. 
In section 3 we will calculate the
supercharges and associated algebra by the use of the Dirac bracket procedure. 
In order to discuss the extended objects, we carefully analyze
the surface terms. In section 4, the boundary conditions for the open
supermembranes on the pp-wave will be considered. Section 5 is
devoted to considerations and discussions. In appendix, our notation 
is summarized.

\section{Supermembrane on Maximally Supersymmetric PP-wave}

We consider the (closed and open) supermembranes
\cite{sezgin,bergshoeff,
dWHN} (for the review, see \cite{T,duff,NH,dW1,dW2,Taylor,DNP}) 
on the eleven-dimensional maximally supersymmetric pp-waves (Kowalski-Glikman
(KG) solution) \cite{KG}. Its metric is given as
\begin{eqnarray}
&&ds^2 = - 2 dx^+ dx^- + G_{++}(dx^+)^2 + \sum_{\mu =1}^9(dx^{\mu})^2\, , \\
& & \qquad G_{++} \equiv - \left[\left(\frac{\mu}{3}\right)^2 
 (x_1^2 + x_2^2 + x_3^2) + \left(\frac{\mu}{6}\right)^2 (x_4^2 + \cdots + 
 x_9^2)\right]\, , \nn  
\end{eqnarray}
where the constant 4-form flux for $+$, 1, 2, 3 directions,    
\begin{equation}
 F_{+123} = \mu\, ,\quad (\mu \neq 0)\label{flux}
\end{equation}
is equipped. 
 
The Lagrangian of supermembrane on the maximally supersymmetric
pp-wave is given as a sum of $\mathcal{L}_0$ and 
Wess-Zumino term $\mathcal{L}_{WZ}$
\footnote{Our notation and
convention are summarized in Appendix.} 
\begin{eqnarray}
&& \mathcal{L} =  \mathcal{L}_0
+ \mathcal{L}_{WZ}\,,\,\,\,
\mathcal{L}_0=
- \sqrt{-g(X,\th)}  \,,
\label{Lag}
\label{start}
\end{eqnarray}
where the induced metric $g_{ij}$ is given by
\begin{equation}
 g_{ij} \; = \; \Pi_i^{\hat{r}} \Pi^{\hat{s}}_j\, \eta_{
\hat{r}\hat{s}}\, , \quad g \;=\; {\rm det}g_{ij}
\end{equation}
and the supervielbein $\Pi^{A}$ and covariant derivative $D_i\th$ for
$\th$
are defined by using vielbein $e^{\hat{r}}_{\hat{\mu}}$ and 
spin connection $\omega^{\hat{r}\hat{s}}$
\begin{eqnarray}
\label{sv1}
 & & \Pi^{\hat{r}} \;=\; d X^{\hat{\mu}}e^{\hat{r}}_{\hat{\mu}} 
- i\bar{\th}\Gamma^{\hat{r}}D\th\, , \\
\label{sv2}
& & \Pi^{\bar{\alpha}} = (D\th)^{\bar{\alpha}} 
\; \equiv \; d\th^{\bar{\alpha}} 
+ e^{\hat{r}}(T_{\hat{r}}^{~\hat{s}\hat{t}\hat{u}\hat{v}}\th)^{\bar{\alpha}}
 F_{\hat{s}\hat{t}\hat{u}\hat{v}} 
- \frac{1}{4}\omega^{\hat{r}\hat{s}}
(\Gamma_{\hat{r}\hat{s}}\th)^{\bar{\alpha}}\,, \\
& & T_{\hat{r}}^{~\hat{s}\hat{t}\hat{u}\hat{v}} \; = \; \frac{1}{288}
(\Gamma_{\hat{r}}^{~\hat{s}\hat{t}\hat{u}\hat{v}} - 8 \delta_{\hat{r}}^{
~[\hat{s}} \Gamma^{\hat{t}\hat{u}\hat{v}]} )\, .
\end{eqnarray}

The maximally supersymmetric 
pp-wave background is achieved by 
taking Penrose limit \cite{OP}
in the $AdS_4 \times S^7$ or $AdS_7 \times S^4$ 
where the supervielbeins are given by \cite{Claus,dW3}
\begin{eqnarray}
 & & \Pi^{\bar{\alpha}} \; = \; \left(
\frac{{\rm sinh}\mathcal{M}}{\mathcal{M}} D\th \right)^{\bar{\al}}\, , \\
& & \Pi^{\hat{r}} = dx^{\hat{\mu}} e^{\hat{r}}_{\hat{\mu}} - i \bar{\th}
\Gamma^{\hat{r}}\left(\frac{2}{\mathcal{M}}{\rm sinh}\frac{\mathcal{M}}{2}
\right)^2 D\th\, ,\\
& &  i\mathcal{M}^2 \; = \; 2(T_{\hat{r}}^{~\hat{s}\hat{t}\hat{u}\hat{v}}\th )
F_{\hat{s}\hat{t}\hat{u}\hat{v}}(\bar{\th}\Gamma^{\hat{r}}) - \frac{1}{288}
(\Gamma_{\hat{r}\hat{s}}\th)[
\bar{\th}(\Gamma^{\hat{r}\hat{s}\hat{t}\hat{u}\hat{v}\hat{w} }F_{\hat{t}\hat{u}\hat{v}\hat{w}} + 24\Gamma_{\hat{t}\hat{u}} F^{\hat{r}\hat{s}\hat{t}\hat{u}} 
)]\, .
\end{eqnarray}
When we take the light-cone gauge
in the Penrose limit, $\mathcal{M}^2 = 0$ is satisfied.
In addition the $D\theta$ becomes a
simple formula
\begin{eqnarray}
\label{2.10}
D\theta = d\th + e^{+} T_{+}^{~+123}\th F_{+123} \, ,
\end{eqnarray}
and supervielbeins are dramatically simplified, though the action 
in the $AdS$ background has non-trivial interaction terms. 
In this gauge we write down the Wess-Zumino term $\mathcal{L}_{WZ}$
\begin{eqnarray}
&&\mathcal{L}_{WZ}=\frac{1}{6}
\epsilon^{ijk}C_{\hat{\mu}\hat{\nu}\hat{\rho}}
\partial_i X^{\hat{\mu}} \partial_j X^{\hat{\nu}} 
\partial_k X^{\hat{\rho}} \nn  \\
&&\qquad   + \frac{i}{2} \epsilon^{ijk}\bar{\th} \Gamma_{\hat{\mu}\hat{\nu}}
D_{i}\th\left(\Pi^{\hat{\mu}}_{j} \Pi^{\hat{\nu}}_{k} 
+ i\Pi^{\hat{\mu}}_{j}\bar{\th}\Gamma^{\hat{\nu}}D_{k}\th 
- \frac{1}{3}\bar{\th}\Gamma^{\hat{\mu}}D_{j}\th \bar{\th}
\Gamma^{\hat{\nu}}D_{k}\th \right)\, .
\end{eqnarray}
Here the supervielbeins on the maximally supersymmetric
 pp-wave are given by Eqs.\,(\ref{sv1}) and (\ref{2.10}). 
$C_{\hat{\mu}\hat{\nu}\hat{\rho}}$ is the $3$-form potential and 
its field strength is described by Eq.(\ref{flux}). 
The above supermembrane action is difficult to analyze directly, 
and  so we shall rewrite Lagrangian (\ref{Lag}) 
following the work \cite{dWHN} in the light-cone gauge
in terms of $SO(9)$ spinor $\psi$ 
\begin{eqnarray}
 w^{-1}\mathcal{L} &=& \frac{1}{2}D_{\tau}X^r D_{\tau}X^r 
- \frac{1}{4}(\{X^r,\,X^s\})^2 - \frac{1}{2}\left(\frac{\mu}{3}\right)^2
\sum_{{\s I}=1}^{3}X_{\s I}^2 - \frac{1}{2}\left(\frac{\mu}{6}\right)^2 
\sum_{{\s I'}=4}^{9}X^2_{\s I'}    \nn \\
& & - \frac{\mu}{6} \sum_{{\s I,J,K}=1}^3\epsilon_{\s IJK}X^{\s K}
\{X^{\s I},\,X^{\s J}\} + i {\psi}^{\s T}\gamma^{r}\{X^r,\,\psi \} 
+ i{\psi}^{\s T} D_{\tau}\psi + i\frac{\mu}{4}\psi^{\s T}\gamma_{\s 123} 
\psi \, .
\label{memb}
\end{eqnarray}
We used a convention $P^{+}_0=1$. 
Here ``$\tau$'' is the time coordinate on the
worldvolume and 
$\{\, , \, \}$ is Lie bracket given by using 
an arbitrary function $w(\sigma)$ of 
worldvolume spatial coordinates $\sigma^a$ ($a=1,2$)
\[
 \{A\,, B\} \; \equiv \; \frac{1}{w}
\epsilon^{ab} \partial_{a} A \partial_{b} B \, ,
\quad (\,a,b = 1,2\,) .
\]
with $\partial_a=\frac{\partial}{\partial \sigma^a}$.
Also this theory has large residual gauge symmetry called 
the area-preserving diffeomorphism (APD) 
and the covariant derivative for this gauge symmetry
is defined by a gauge connection $\omega$ 
\begin{equation}
 D_{\tau} X^r \;\equiv\; \pd_{\tau}X^r - \{\omega,\,X^r \}\, .
\end{equation}
In this model, if we replace the variables in the Lagrangian (\ref{memb})
following the rule with
\begin{eqnarray}
 X(\xi^{i}) & \longrightarrow & X(\tau) \nn \\
 \psi(\xi^i) & \longrightarrow & \psi(\tau) \nn \\
 \int\!d^2\sigma\, w(\sigma) & \longrightarrow & {\rm Tr} \nn \\
 \{~ , ~\} & \longrightarrow & -i[~ , ~ ]\, , \nn  
\end{eqnarray}
we can obtain the BMN matrix model, 
starting from the Lagrangian for supermembrane on the maximally
supersymmetric pp-wave.

We have taken the light-cone gauge 
and so original symmetries are not seen manifestly 
but the Lagrangian (\ref{memb}) still has residual supersymmetries, 
\begin{eqnarray}
 & & \del_{\epsilon} X^r \;=\; 2\psi^{\s T}\gamma^r \epsilon(\tau)\, , 
\quad \del_{\epsilon} \omega \;=\; 2\psi^{\s T} \epsilon(\tau)\, , \nn \\
\label{linear}
& &  \del_{\epsilon}\psi \;=\; - i D_{\tau} X^r \gamma_r \epsilon(\tau) 
\,+\, \frac{i}{2}\{X^r,\, X^s \}\gamma_{rs}\epsilon(\tau)  \\
& & \,\qquad + \, \frac{\mu}{3} i 
\sum_{I=1}^{3}X^{\s I} \gamma_{\s I}\gamma_{\s 123}\epsilon(\tau) 
\,-\, \frac{\mu}{6} i\sum_{I'=4}^{9} X^{\s I'}\gamma_{\s I'}
\gamma_{\s 123}
\epsilon(\tau)\, ,\nn \\
& & \epsilon(\tau) \;=\; \exp\left(\frac{\mu}{12}\gamma_{\s 123}\tau \right)
\epsilon_0\, \quad (\,\epsilon_0:\,{\rm constant~spinor}). \nn
\end{eqnarray}
These transformation rules are 16 linearly-realized supersymmetries on
the maximally supersymmetric pp-wave. 
In taking the limit, $\mu \rightarrow 0$,  
we recover the supersymmetry   
transformations on the flat space. 
In the context of the eleven dimensional supersymmetry, this corresponds to 
the dynamical supersymmetry.
The Lagrangian (\ref{memb}) has other 16 nonlinearly realized
supersymmetries,
\begin{eqnarray}
 & & \del_{\eta}X^r \;=\; 0\,,\quad \del_{\eta}\omega \;=\; 0\, , \nn \\
\label{nonlinear}
 & & \del_{\eta}\psi \;=\; \eta(\tau)\, ,  \\
 & & \eta(\tau) \;=\; \exp\left( - \frac{\mu}{4}\gamma_{\s 123} \tau \right)
\eta_0\, , \quad (\,\eta_0:\,{\rm constant~spinor}). \nn
\end{eqnarray} 
It corresponds to the kinematical supersymmetry in the 
eleven dimensional theory.

\section{Supercharge Algebra from the Supermembrane} 

To begin, we derive supercharges for the supersymmetries 
(\ref{linear}) and (\ref{nonlinear}), and then study 
associated superalgebra
by the use of the Dirac bracket. 
We discuss the extended objects on the pp-wave 
from the viewpoint of the 
central charges of the superalgebra. 

Supercharges $Q^+$ and $Q^-$ 
of the linearly and non-linearly realized supersymmetries, respectively,  
are obtained as Noether charges
\begin{eqnarray}
&& Q^+ = \int \! d^2\sigma\, w 
\Bigg[ -2 \e^{-\frac{\mu}{12}\gamma_{\s 123}\tau} 
\Big( DX^r \gamma_r\psi + \frac{1}{2}\{X^r,\,X^s\}\gamma_{rs}\psi \nn \\
& &\qquad  + \frac{\mu}{3}\sum_{\s I=1}^3
X^{\s I}\gamma_{\s I}\gamma_{\s 123}\psi 
+ \frac{\mu}{6}\sum_{\s I'=4}^9 X^{\s I'}\gamma_{\s I'}\gamma_{\s 123}\psi 
\Big)\Bigg]\,, 
\label{ch-l}
\\
&&Q^- = \int \! d^2\sigma\, 
w\left[-2i \e^{\frac{\mu}{4}\gamma_{\s 123}\tau}\psi
\right]\, \nn \\
&&\qquad = -2i \e^{\frac{\mu}{4}\gamma_{\s 123}\tau}\psi_0\, ,
\label{ch-nl}
\end{eqnarray}
where $\psi_0$ is the zero-mode of $\psi$ and we have used the
normalization with
$\int \! d^2\sigma\, w(\sigma)= 1$. 

Next we shall calculate the superalgebra satisfied by (\ref{ch-l})
and (\ref{ch-nl}). 
The supermembrane theory contains the fermionic field $\psi^{\al}$ 
and this leads to 
the second class constraint $\Xi_{\al}\approx 0$ for the theory 
\begin{equation}
 \Xi_{\al} \;=\; S_{\al} - i w\psi^{\s T}_{\al} \; \approx\; 0
\,,\quad \left(S_{\al} \equiv 
\frac{\partial \mathcal{L}}{\partial_{\rm\s R}(\partial_0\psi^{\al}) } 
= iw\psi^{\s T}_{\al} \right)\, .
\label{con}
\end{equation} 
The fermionic field is the SO(9) spinor with 16 components.
We have to deal properly with this second class constraint 
by the use of the Dirac bracket. The calculation of the Dirac bracket 
needs only a constraint matrix $C_{\al\beta}$ 
\begin{equation}
 C_{\al\beta} \;\equiv\; \{\Xi_{\al}(\sigma),\,
\Xi_{\beta}(\sigma')\}_{\rm\s PB} 
= -2i w\del_{\al\beta} \del^{(2)}(\sigma - \sigma')\, ,
\end{equation} 
and its inverse matrix $(C^{-1})_{\al\beta}$ is given by 
\begin{equation}
 (C^{-1})_{\al\beta} \;=\; \frac{i}{2w} \del_{\al\beta}\del^{(2)}
(\sigma - \sigma'). 
\end{equation}
By the use of the matrix
$(C^{-1})_{\al\beta}$, we can introduce the Dirac bracket $\{\,,\,\}_{DB}$ 
in terms of the Poisson bracket $\{\,,\,\}_{PB}$
\begin{eqnarray}
 \{F,\,G \}_{\rm\s DB} & \equiv & \{F,\,G\}_{\rm\s PB} 
- \{F,\,\Xi_{\al}\}_{\rm\s PB}(C^{-1})^{\al\beta}
\{\Xi_{\beta},\, G\}_{\rm\s PB}, \nn \\
&=& \{F,\,G\}_{\rm\s PB} - \frac{i}{2w}\{F,\,\Xi_{\al}\}_{\rm\s PB}
\{\Xi^{\al},\,G\}_{\rm\s PB}.
\end{eqnarray}
Thus, we can define the commutation relations on the bosonic and
fermionic fields with their canonical momenta  
\begin{eqnarray}
\label{DB-boson}
  &&\{X^{r}(\sigma),\,P_{s}(\sigma')\}_{\rm\s DB} 
= \del^{r}_{s}\del^{(2)}(\sigma - \sigma')\, , \\ 
&&\{\psi_{\al}(\sigma),\,S_{\beta}(\sigma')\}_{\rm\s DB} 
= \frac{1}{2}\del_{\al\beta}\del^{(2)}(\sigma - \sigma')\,,
\label{DB-fer}\\
&& \qquad P_r=wD_{\tau}X_r\,.\nonumber
\end{eqnarray}
The commutation relations (\ref{DB-boson}),(\ref{DB-fer}) are rewritten in terms of
$\psi^{\s T}$ and $D_{\tau}X_s$ 
\begin{eqnarray}
&&\{X^r,D_{\tau}X_s\}_{\rm\s DB}=\frac{1}{w}\delta_{s}^{r}\delta^{(2)}(\sigma-\sigma')\,,\label{DB-bo}\\
&& \{\psi_{\al}(\sigma),\,\psi^{\s T}(\sigma') \}_{\rm\s DB} = 
-\frac{i}{2w}\del_{\al\beta}\del^{(2)}
(\sigma - \sigma')\,.
\label{DB-fermion}
\end{eqnarray}
The superalgebra is calculated by the use of the Dirac bracket
(\ref{DB-boson}),(\ref{DB-fer}),(\ref{DB-bo}),(\ref{DB-fermion})
and we obtain the results 
\begin{eqnarray}
& & i\left\{\frac{1}{\sqrt{2}}Q_{\al}^-,\,\frac{1}{\sqrt{2}}
(Q^-)_{\beta}^{\s T}\right\}_{\rm\s DB} \;=\; 
 - \del_{\al\beta}\, , \\
& & i\left\{\frac{1}{\sqrt{2}}Q_{\al}^+,\,\frac{1}{\sqrt{2}}
(Q^-)_{\beta}^{\s T}\right\}_{\rm\s DB} \;=\; 
  i\sum_{{\s I}=1}^{3} \left[ \left(P_0^{\s I} + \frac{\mu}{3}
X^{\s I}_0 \gamma_{\s 123}\right)\gamma_{\s I} 
\e^{-\frac{\mu}{3}\gamma_{\s 123}\tau}
\right]_{\al\beta}  \\
& & \qquad  +  i\sum_{{\s I'}=4}^{9} \left[ \left(
P_0^{\s I'} - \frac{\mu}{6}X^{\s I'}_0
\gamma_{\s 123} \right)\gamma_{\s I'} 
\e^{-\frac{\mu}{6}\gamma_{\s 123}\tau}
\right]_{\al\beta} - i\sum_{{\s I,J}=1}^3 
\int \! d^2\sigma\, \partial_{a}S^a_{\s IJ}
\left(\gamma^{\s IJ}\e^{-\frac{\mu}{3}\gamma_{\s 123}}\right)_{\al\beta} 
\nn \\ 
& & \qquad - i\sum_{{\s I',J'}=4}^9 
\int \! d^2\sigma\, \partial_{a}S^a_{\s I'J'}
\left(\gamma^{\s I'J'}\e^{-\frac{\mu}{3}\gamma_{\s 123}}\right)_{\al\beta} 
- 2i\sum_{{\s I}=1}^3\sum_{{\s I'}=4}^9 
\int \! d^2\sigma\, \partial_{a}S^a_{\s II'}
\left(\gamma^{\s II'}\e^{-\frac{\mu}{6}\gamma_{\s 123}}\right)_{\al\beta}
\, , \nn 
\end{eqnarray}
\begin{eqnarray}
& & i\left\{\frac{1}{\sqrt{2}}Q_{\al}^+,\,\frac{1}{\sqrt{2}}
(Q^+)_{\beta}^{\s T}\right\}_{\rm\s DB} \;=\; 
2H \delta_{\alpha\beta}  \\
& & \qquad +  \frac{\mu}{3}\sum_{{\s I,J}=1}^3 M^{\s IJ}_0 
\left(\gamma_{\s IJ}\gamma_{\s 123}\right)_{\al\beta} 
- \frac{\mu}{6}\sum_{{\s I',J'}=4}^9 
M_0^{\s I'J'}
\left(\gamma_{\s I'J'}\gamma_{\s 123}\right)_{\al\beta} \nn \\
& & \qquad - 2 \sum_{{\s I}=1}^3
\int\! d^2\sigma\, \varphi X_{\s I}(\gamma^{\s I})_{\al\beta}
-2\sum_{{\s I'=4}}^9\int\!d^2\sigma\,\varphi X_{\s I'}
\left(\gamma^{{\s I'}}\e^{\frac{\mu}{6}\gamma_{\s 123}\tau}\right)_{\al\beta} 
\nn \\
& & \qquad + 2 \sum_{{\s I}=1}^3
\int\! d^2\sigma\, \partial_a S^{a}_{\s I}  (\gamma^{\s I})_{\al\beta}  
+ 2 \sum_{{\s I'}=4}^9
\int\!d^2\sigma\, \partial_a S^{a}_{\s I'}
 \left(\gamma_{\s I'}\e^{\frac{\mu}{6}\gamma_{\s 123}\tau}
\right)_{\al\beta}
\nn \\       
& & \qquad + 2\sum_{{\s I,J}=1}^3\sum_{{\s I',J'}=4}^9
\int\!d^2\sigma\,
\partial_{a}S^{a}_{IJI'J'} \left(\gamma^{\s IJI'J'}\right)_{\al\beta} 
+ 2\!\!\! \sum_{{\s I',J',K',L'}=4}^9\int\!d^2\sigma\,
\partial_{a}S^{a}_{I'J'K'L'} \left(\gamma^{\s I'J'K'L'}\right)_{\al\beta} 
\nn \\
& & \qquad  + 2 \sum_{{\s I,J,K}=1}^3\sum_{{\s I'}=4}^9
\int\!d^2\sigma\,
\partial_{a}S^{a}_{IJKI'} \left(\gamma^{\s IJKI'}\e^{\frac{\mu}{6}
\gamma_{\s 123}\tau}\right)_{\al\beta}  \nn \\
& & \qquad 
+ 2 \sum_{{\s I}=1}^3\sum_{{\s I',J',K'}=4}^9
\int\!d^2\sigma\,
\partial_{a}S^{a}_{II'J'K'}  \left(\gamma^{\s II'J'K'}\e^{\frac{\mu}{6}\gamma_{\s 123}\tau}
\right)_{\al\beta} 
\nn \\
& & \qquad + 2\mu\sum_{{\s J,K}=1}^3\sum_{{\s I',J'}=4}^9
\int\!d^2\sigma\,
\partial_{a}U_{\s JKI'J'}^{a}\left(\gamma^{\s JK}\gamma^{\s I'J'}\right)_{\al\beta} \nn \\ 
& & \qquad 
+ 2\mu \sum_{{\s I'}=4}^9\int\!d^2\sigma\,
\partial_{a}U_{I'}^{a}\left(\gamma^{\s I'}\gamma_{\s 123}\e^{\frac{\mu}{6}\gamma^{\s 123}\tau}\right)_{\al\beta} 
\,. \nn
\end{eqnarray}
Here $M^{\s IJ}$ and $M^{\s I'J'}$ are defined by 
\begin{eqnarray}
& &  M^{\s IJ} \;=\; X^{\s I}P^{\s J} - P^{\s I}X^{\s J} - 
\frac{1}{2}S^{\s T}\gamma^{\s IJ}\psi\,, \\
& &  M^{\s I'J'} \;=\; X^{\s I'}P^{\s J'} - P^{\s I'}X^{\s J'} - 
\frac{1}{2}S^{\s T}\gamma^{\s I'J'}\psi\,, 
\end{eqnarray} 
and the $SO(3)\times SO(6)$ Lorentz generators 
$M^{\s IJ}_0$ and $M^{\s I'J'}_0$ are given as
\begin{eqnarray}
& & M^{\s IJ}_0 \;\equiv\; \int\!d^2\sigma\,M^{\s IJ}\, , \\
& & M^{\s I'J'}_0 \;\equiv\; \int\!d^2\sigma\,M^{\s I'J'}\, .
\end{eqnarray}
They satisfy the $SO(3)\times SO(6)$ Lorentz algebra, 
\begin{eqnarray}
& & \left\{M_0^{\s IJ},\, M_0^{\s KL} \right\}_{\rm DB} \;=\;
\del^{IK}M_0^{\s JL} - \del^{\s IL}M_0^{\s JK} - \del^{\s JK}M_0^{\s IL} 
+ \del^{\s JL}M_0^{\s IK}\,, \\
& & \left\{M_0^{\s I'J'},\, M_0^{\s K'L'} \right\}_{\rm DB} \;=\;
\del^{I'K'}M_0^{\s J'L'} - \del^{\s I'L'}M_0^{\s J'K'} - \del^{\s J'K'}
M_0^{\s I'L'} 
+ \del^{\s J'L'}M_0^{\s I'K'}\,.
\end{eqnarray}
The zero-modes of $P^r (= w D_{\tau}X^{r})$ and $X^r$ are 
written by 
\begin{eqnarray}
 P_0^r &\equiv& \int\! d^2\sigma\, w D_{\tau}X^{r}\, , 
\,\,\,X^{\s r}_0=\int\!d^2\sigma\,wX^{\s r}\,. 
\end{eqnarray}
Also, the Hamiltonian H is expressed as 
\begin{eqnarray}
& & H \;=\; \int\!d^2\sigma\,w\Bigg[\frac{1}{2}\left(\frac{P^r}{w}\right)^2
+ \frac{1}{4} \{X^r,X^s\}^2
+ \frac{1}{2} \left(\frac{\mu}{3}\right)^2\sum_{{\s I}=1}^{3}(X^{\s I})^2
+ \frac{1}{2} \left(\frac{\mu}{6}\right)^2\sum_{{\s I'}=4}^{9}(X^{\s I'})^2
\nn\\
& &\qquad +\frac{\mu}{6}\sum_{{\s I,J,K}=1}^{3}\epsilon_{\s IJK}
X^{\s K} \{X^{\s I},X^{\s J} \} - w^{-1}\frac{\mu}{4}S^{\s T}\gamma_{\s 123}
\psi -w^{-1}S^{\s T}\gamma_{r}\{X^r,\psi\}\Bigg]\,.
\end{eqnarray}
Other quantities in the above algebra are defined by
\begin{eqnarray}
 S^a_{rs} &\equiv& -\frac{1}{2}\epsilon^{ab}X^{[r}\partial_b X^{s]}\, , \\
 \varphi &\equiv& w \{w^{-1}P^r,\,X^r\} + iw \{\psi^{\s T},\,\psi\}\, , \\
 S^a_r &\equiv& \epsilon^{ab}
\left(w^{-1}X_rP_s\partial_{b}X^s + X_r i\psi^{\s T}\partial_{b}\psi
 + \frac{3}{8}iX^s\partial_b(\psi^{\s T}\gamma_{rs}\psi)\right)\, , \nn \\ 
 S^a_{rstu} &\equiv& \frac{i}{48}\epsilon^{ab}X_{[r}\partial_b \left(
\psi^{\s T}\gamma_{stu]}\psi\right)
\, ,\\
 U_{\s JKI'J'}^a &\equiv& - \frac{1}{6}\sum_{I=1}^3
\epsilon_{\s IJK}\epsilon^{ab} X^{\s I'}\partial_{b}(
X^{\s I}X^{\s J'})\,, \\ 
 U_{I'}^a &\equiv& - \frac{1}{2}
\epsilon^{ab} X^{\s I'}\partial_{b}
\left[\frac{1}{3}\sum_{{\s I=1}}^3(X^{\s I})^2 
- \frac{1}{6}
\sum_{{\s J'}=4}^9
(X^{J'})^2 
\right]\,. \nn 
\end{eqnarray}
The above superalgebra (other than the central charges) 
completely agrees with that of the BMN matrix model \cite{Malda} 
\footnote{We can absorb the factor $1/\sqrt{2}$ in front of the
supercharge in the definition of the fermions $\psi$.}. 
Also, in the $\mu\rightarrow 0$ limit, the above algebra realize 
the superalgebra of the supermembrane in the flat
space given in \cite{dWHN}.  

Also, the above superalgebra  
includes some central charges. These charges indicate the 
existence of extended objects in supermembrane theories 
on the maximally supersymmetric pp-wave. First, the charges $S^a_{rs}$ 
and $S_r^a$ correspond 
to the transverse M2-brane (D2-brane in type IIA string theory) 
and longitudinal M2-brane (fundamental string in type
IIA string theory), respectively. 
Next, $S^a_{rstu}$ corresponds to the longitudinal M5-brane
charge (D4-brane in type IIA string theory). As is well-known, these 
charges have appeared in the supermembrane theory on the 
flat eleven-dimensional Minkowski space. 
In addition, in our supermembrane theory the 
superalgebra includes the additional central charges, 
$U_{\s JKI'J'}^a$ and $U_{\s I'}^a$. We do not properly confirm 
the physical interpretation of these extra extended objects 
only living on the pp-wave. These might be related to 
the fuzzy membrane and  
giant graviton discussed in \cite{Malda}, or another new extended 
object due to a certain kind of the Myers effects on the pp-wave
\cite{Myers}.

\section{Open Supermembrane on PP-wave}

In the case of the open membrane, which has the boundary on the worldvolume 
toward the spatial directions $\sigma^1$ and $\sigma^2$, the surface
terms do not vanish automatically.  Thus we must properly 
treat the total derivative terms under the variation of 
the above supersymmetry transformations, and consider the boundary
conditions in order for the surface terms to vanish. 
Let us recall that the membrane $p$-branes are allowed for $p=1,\,5,\,9$ 
in the flat background due to 
the boundary conditions \cite{EMM}. M5-brane corresponds to $p=5$. The
case of $p=9$ is related to ``the end of the world'' in
Ho$\breve{\rm r}$ava-Witten's works \cite{HW}. 

In our pp-wave case, we obtain 
the total derivative terms for the linear supersymmetry (\ref{linear}) 
explicitly
\begin{eqnarray}
& & w\left\{X^r,\, 
D_{\tau}X^s \psi^{\s T}\gamma_s\gamma_r\epsilon(\tau) + \frac{1}{2}\{X^s,\, X^t\}\psi^{\s T}\gamma_{st}\gamma_{r}\epsilon(\tau) \right\} \nn \\
& & - \frac{\mu}{3}\sum_{{\s I,J}=1}^3 w\{X^{\s I},\, X^{\s J}\psi^{\s T} 
\gamma_{\s I}\gamma_{\s J}\gamma_{\s 123} \epsilon(\tau) \} 
- \frac{\mu}{6} \sum_{{\s I',J'} = 4}^9 
w\{X^{\s I'},\, X^{\s J'}\psi^{\s T}\gamma_{\s I'}
\gamma_{\s J'}\gamma_{\s 123}\epsilon(\tau) \} \nn \\
& & + \frac{\mu}{3}\sum_{{\s I}=1}^3\sum_{{\s I'}=4}^9 
w\{X^{\s I'},\, X^{\s I} 
\psi^{\s T} \gamma_{\s I' I}\gamma_{\s 123}\epsilon(\tau)\} - \frac{\mu}{6}
\sum_{{\s I}=1}^3\sum_{{\s I'}=4}^9 w\{X^{\s I},\, X^{\s I'}\psi^{\s T}
\gamma_{\s I I'}\gamma_{\s 123}\epsilon(\tau) \}\, ,
\label{td-eps}
\end{eqnarray}
and for the nonlinear supersymmetry (\ref{nonlinear}), we can calculate the 
corresponding term 
\begin{eqnarray}
w\{X^r,\, i\eta \gamma_r \psi\}\,.
\label{td-eta}
\end{eqnarray} 
This surface term for the nonlinear supersymmetry has the same form as 
the flat space case. However, some additional terms proportional to $\mu$ 
appear for the linear supersymmetry 
in addition to the surface terms in the flat background. The variations of the
action under the linear and nonlinear supersymmetry transformations can
be written as 
\begin{eqnarray}
\del S &=& \del_{\epsilon}S + \del_{\epsilon}^{\s (\mu)}S + \del_{\eta}S 
\, , \nn \\
\label{flat-ep} 
\del_{\epsilon}S &=& - \int \!\!d\tau\!\! \int_{\partial\Sigma}\!\!\! d\xi\,
 \Bigg[
\partial_{\bf t} X^r \cdot \left(D_{\tau}X^s\psi^{\s T}\gamma_s\gamma_r \epsilon(\tau) 
+ \frac{1}{2}\{X^s,\,X^t\}\psi^{\s T}\gamma_{st}\gamma_r\epsilon(\tau) \right)
\Bigg]\,,  \\
\label{pp}
\del_{\epsilon}^{\s (\mu)}S &=& 
 - \int \!\!d\tau\!\! \int_{\partial\Sigma}\!\!\! d\xi\, \Bigg[
- \frac{\mu}{3}\sum_{{\s I,J}=1}^3\partial_{\bf t} X^{\s I} \cdot X^{\s J} \psi^{\s T}\gamma_{\s I}\gamma_{\s J}\gamma_{\s 123}\epsilon(\tau) 
- \frac{\mu}{6}\sum_{{\s I',J'}=4}^{9}\partial_{\bf t}X^{\s I'}\cdot X^{\s J'}
\psi^{\s T}\gamma_{\s I'}\gamma_{\s J'}\gamma_{\s 123}\epsilon(\tau) \nn \\
& & + \frac{\mu}{3}\sum_{{\s I}=1}^3\sum_{{\s I'}=4}^9 \partial_{\bf t}
X^{\s I'} \cdot X^{\s I} \psi^{\s T}\gamma_{\s I' I} \gamma_{\s 123}
\epsilon(\tau) - \frac{\mu}{6} \sum_{{\s I}=1}^3\sum_{{\s I'}=4}^9 
\partial_{\bf t}X^{\s I}\cdot X^{\s I'}\psi^{\s T}\gamma_{\s I I'}
\gamma_{\s 123} \epsilon(\tau)\,  
\Bigg] \, , \\
\label{flat-eta}
\del_{\eta}S &=& -i \int \!\! d\tau\!\! \int_{\partial\Sigma}\!\!\! d\xi\, 
\partial_{\bf t}
X^r \cdot 
\eta(\tau) \gamma_{r}\psi
\, , 
\end{eqnarray} 
 where $\partial\Sigma$ is the boundary of the open supermembrane
 worldvolume and 
$\xi$ is the coordinate for the tangent direction of the boundary. 
Note that the tangential derivative $\partial_{\bf t}$ and normal derivative 
$\partial_{\bf n}$ on the
boundary are  defined by 
\begin{eqnarray}
 \partial_{\bf t}X^r &\equiv& \epsilon^{ab}n_a\partial_b X^r\,, \\
 \partial_{\bf n}X^r &\equiv& n^a\partial_a X^r \, .
\end{eqnarray}
Here $n^a$ is the unit vector toward the normal direction on the
boundary. We would like to consider the $p$-dimensional hypersurface 
(membrane $p$-brane) on which supermembranes can end, 
and investigate the 
condition that such a surface can exist.  
First, by following the discussion of the $p$-brane in string theory, 
the boundary
conditions for our membrane are classified
\begin{eqnarray}
&& {\rm Neumann}\,:\quad \partial_{\bf n} X^{\overline{m}} \;=\; 0\,,
\quad (\overline{m}=0,10\,{\rm and~some~}p-1{\rm ~coordinates} ) \\
&& {\rm Dirichlet}\,:\quad \partial_{\bf t} X^{\underline{m}} \;=\; 0\,,\quad
(\underline{m}= {\rm other~}10-p~{\rm coordinates}) \,.
\end{eqnarray}
By applying these boundary conditions to (\ref{flat-ep})
and (\ref{flat-eta}), the constraints 
\begin{eqnarray}
 \eta_{0}^{\s T}\gamma_{\s\overline{m}}\psi = \epsilon_0^{\s T}\gamma_{\s\overline{m}}\gamma_{\s\overline{n}}\psi = \epsilon_0^{\s T}\gamma_{\s\overline{m}}\gamma_{\s\underline{n}}\gamma_{\s\overline{n}}\psi = 0 \, , 
\label{159}
\end{eqnarray}
can be obtained.
These are the same conditions as the flat case
 and (\ref{159}) 
leads us to the well-known results $p=1,\,5$ and 9. 
However, in the pp-wave case we also need to take account of  
the constraints coming from the additional surface terms (\ref{pp}). 

Here, let us define the following operators
\begin{equation}
 P_{\pm} \;\equiv\; \frac{1}{2}\left(1 \pm \gamma^{\underline{m}_1}
\gamma^{\underline{m}_2}\ldots\gamma^{\underline{m}_{10-p}} \right).
\end{equation}  
These are the projection operators if and only if 
$\frac{1}{2}p(p+1)$ is odd. Thus, the value of $p$ is 
limited to $p=1,\,2,\,5,\,6$ and 9. 
The requirement that boundary term should vanish leads to 
the constraints Eq.\,(\ref{159}), and so it provides a 
further restriction for the value
of $p$. If we assume that 1/2 BPS boundary hypersurface, then 
the condition 
\begin{equation}
 P_- \psi \;=\; 0\,, 
\end{equation}
is in our hand. Then we can write $\psi$ as 
$\psi = P_+ {\psi}$. To begin, from the second equation in (\ref{159}), 
$P_+ \epsilon_0 =0$ is followed. Next, we can read off 
from the third equation in
(\ref{159}) that $9-p$ should be even. As a result, $p=1,\,5$ and 9 are
allowed in the flat case for the boundary hypersurface. 
However, the story does not end because the additional boundary terms 
exist in the case of the pp-wave. 
We can easily check whether the additional surface terms (\ref{pp})
vanish or not in each $p=1,\,5,\,9$ case. In the  $p=1$ case we can
immediately see that the additional terms (\ref{pp}) vanish. 
Here, it can be seen from the constraints (\ref{159}) that only the
even number of gamma matrices with Neumann indices $\overline{m}$'s 
and arbitrary number of gamma matrices with Dirichlet indices
$\underline{m}$'s are allowed to
appear between $\epsilon^{\s T}_0$ and $\psi$. Equivalently, odd number
of gamma matrices with Neumann indices $\overline{m}$'s cannot appear 
between $\epsilon^{\s T}_0$ and $\psi$. However, it is found 
from the expression (\ref{pp}) that 
such a condition cannot be satisfied in the cases $p=5$ and $9$
because there are inevitably several terms 
including odd number of Neumann components. 
In conclusion, only $p=1$ is allowed for membrane $p$-brane on the
pp-wave, and $p=5,\,9$ membrane $p$-brane cannot exist.  
   
This result would be also plausible from the viewpoint of 
the chirality matrix \cite{BB}. It is because that the flux is
turned on the $1,\,2,\,3$ directions on the pp-wave, 
and so the $SO(4)$ and $SO(8)$ chirality,  
which is important for $p=5$ and $p=9$ cases, cannot be respected. 
The reason that $p=1$ case is allowed is unknown, since 
what $p=1$ means physically has not been well understood.

In the above discussion, we have assumed that the 1/2 BPS boundary 
hypersurface, that is, the flat boundary
hypersurface. However, it might be clear that such flat boundaries 
cannot exist, because the pp-wave background is curved. 
Possibly, the curved hypersurface as discussed in \cite{bak} 
might become the boundary of the supermembrane. 
But, we do not know how to treat such curved boundaries, and do not 
discuss the case here.

\section{Conclusions and Discussions}

In this paper, we have studied the supercharges and its associated algebra. 
In particular, by treating the surface terms carefully, its central
extension has been derived. The superalgebra apart from the central
charges completely agrees with that of the BMN matrix model. 
The central charges obtained in our derivation realize the 
flat space results in the $\mu \rightarrow 0$, 
and also include some additional ones. 
We do not confirm the physical interpretation of 
the additional central charges. These seem to indicate the extra 
extended objects coming from a kind of the Myers effect on the 
pp-wave background.

Moreover, we have discussed the boundary conditions of the open supermembrane 
on the maximally supersymmetric pp-wave background. It is well-known 
that the membrane $p$-branes in the flat space are allowed to 
exist only for $p=1,\,5$ and 9. In our case on the pp-wave,  
more strict constraints for such 
hypersurfaces arise, and so only the value $p=1$ is allowed. 
In our discussion, we have not included the 2-form which can couple to 
the boundary hypersurface. It might be possible by turning on the
2-form on the boundary that $5$- and 9-dimensional 
hypersurfaces exist as the boundaries of the open supermembranes 
on the pp-wave. 

In this paper, we have used the $SO(9)$ formulation for the simplicity, 
but it is also interesting to work in the $SO(10,1)$ 
covariant formulation, where 
the nature of longitudinal components are clear and more definite 
considerations would be possible. 
This is an interesting future work.  \\

\noindent 
{\bf\large Acknowledgement}

The work of K.S. is supported in part by the Grant-in-Aid from the 
Ministry of Education, Science, Sports and Culture of Japan 
($\sharp$ 14740115). 

\newpage

\appendix 

\noindent
{\bf\large Appendix}

In this appendix, we summarize several notations used in the paper.

\subsection*{Notation}

We consider supermembrane in eleven dimensional curved spacetime and use a notation of 
supercoordinates ($D=11$):
\begin{eqnarray}
X^M = (X^{\hat{\mu}},\th^{\al}), \qquad\hat{\mu} = 
(+,-,\mu),~\mu = 1,\ldots, D-2. \nonumber
\end{eqnarray}
The background metric is expressed as $G_{MN}$.

In the Lorentz frame, we also use the coordinates $(D=11)$:
\begin{eqnarray}
X^A = (X^{\hat{r}},\th^{\bar{\al}}), \qquad \hat{r}=
(+,-,r),~r=1,\ldots,D-2.\nonumber 
\end{eqnarray}
The metric is flat and is described by $\eta_{AB}$. 
In these notations, we introduced a set of light-cone coordinates
$
 X^{\pm} \equiv \frac{1}{\sqrt{2}}(X^0 \pm X^{D-1}) 
$.

The membrane has three-dimensional worldvolume and its coordinates
are 
parameterized by
$\xi^i=(\tau, \sigma^a),~a=1,\,2$, 
and its metric is given by $g_{ij}$.

Next we shall summarize the $SO$(10,1) gamma matrices ($D=11$):
\begin{eqnarray}
 & & \{\Gamma^{\hat{\mu}},\,\Gamma^{\hat{\nu}} \} = 2G^{\hat{\mu}\hat{\nu}}, 
\quad \{\Gamma^{\hat{r}},\,\Gamma^{\hat{s}}\} = 2\eta^{\hat{r}\hat{s}}, \nn \\
& & \Gamma^{\hat{\mu}} \equiv  e^{\hat{\mu}}_{\hat{r}}\Gamma^{\hat{r}}, \quad 
\Gamma^{\hat{r}} \equiv e^{\hat{r}}_{\hat{\mu}}\Gamma^{\hat{\mu}}\, ,  
\nn \\
 & & \Gamma^{\mu} = \gamma^{\mu} \otimes \sigma_3 = 
\begin{pmatrix}
\gamma^{\mu} & 0 \\
0 &  - \gamma^{\mu}
\end{pmatrix}
\, , \quad {\rm (real~symmetric)} \nn \\
& & \Gamma^{0} = 1\otimes i\sigma_2 =
\begin{pmatrix}
0 & - I_{16} \\
I_{16} & 0  
\end{pmatrix}
\, , \quad {\rm (real~skew-symmetric)} \nn \\
& & \Gamma^{D-1} = 1 \otimes \sigma_1 = 
\begin{pmatrix}
0 & I_{16} \\
I_{16} &  0  
\end{pmatrix}
\, , \quad {\rm (real~symmetric)} \nn 
\end{eqnarray}
\begin{eqnarray}
\Gamma^{\pm} \equiv \frac{1}{\sqrt{2}}\left(\Gamma^0 \pm \Gamma^{D-1}\right),
 \quad \{ \Gamma^+,\, \Gamma^- \} = -2 I_{32}\, ,
\nn \\
 \Gamma^+ = \sqrt{2}
\begin{pmatrix}
0 & 0 \\
I_{16} & 0
\end{pmatrix}
, \quad \Gamma^- = \sqrt{2}
\begin{pmatrix}
0 & - I_{16} \\
0 & 0 
\end{pmatrix}
\, . \nn 
\end{eqnarray}

We take the light-cone gauge and decompose the 
32 component $SO$(10,1) spinor $\theta$ in terms of $SO$(9) spinor $\psi$
with 16 components
\begin{eqnarray}
& &  X^{+} = \tau, \quad \Gamma^{+}\th = 0,\quad (\bar{\th}\Gamma^+ = 0)\, ,
 \nn \\
& &  \Longrightarrow \; \th = \frac{1}{2^{1/4} w} 
\dbinom{0}{\psi}\, ,
\nn \\
& & \bar{\th} = \th^{\s T}(- \Gamma^0) = -\frac{1}{2^{1/4}w}
\left(\psi,\, 0\right)\, . \nn 
\end{eqnarray}
In the light-cone gauge, there are several useful identities
\begin{eqnarray}
& & \bar{\th}\Gamma^{\hat{r}}\pd_i\th = 0\, ,\quad 
({\rm for}~\hat{r}\neq -)\, , \nn \\
& & \bar{\th}\Gamma_{rs}\pd_i\th = 0\, , \nn \\
& & \bar{\th}\Gamma^{+r}\pd_i\th = 0\, , \nn \\
& & \bar{\th}\Gamma^{+-}\pd_i\th = 0\, . \nn
\end{eqnarray}
In the pp-wave background, the vielbein is calculated as
\begin{eqnarray}
 & & e^{\hat{r}}_{\hat{\mu}}: \qquad\, 
 e^{+}_{+} = e^{-}_{-} = 1, \quad e^{+}_{-} = 0, \quad e^{-}_{+} = - \frac{1}{4}G_{++}, \quad e^r_{\mu} = \del^r_{\mu}\, , \nn \\
 & & e^{\hat{\mu}}_{\hat{r}}: \qquad\, e^{+}_{+} = e^{-}_{-} = 1, \quad e^{+}_{-} = 0, \quad e^{-}_{+} = 
 \frac{1}{4}G_{++}, \quad e_r^{\mu} = \del_r^{\mu}\, , \nn \\
& & e^{\hat{\mu}\hat{r}}: \qquad\! e^{++} = e^{+r} = e^{-r} = e^{\mu +} = e^{\mu -} = 0\, , \nn \\
& & \qquad \qquad e^{+-} = e^{-+} = - 1,\quad e^{--} = - 
\frac{1}{4}G_{++}, \quad e^{\mu r} = \del^{\mu r}\, , \nn \\
& & e_{\hat{\mu} \hat{r}}: \qquad\! e_{\mu +} = e_{\mu -} = e_{--} = e_{+r} = 
e_{-r} = 0\, , \nn \\ 
& & \qquad \qquad e_{+-} = e_{-+} = -1, \quad e_{++} = + \frac{1}{4}G_{++}, 
\quad e_{\mu \hat{r}} = \del_{\mu r}\,, \nn 
\end{eqnarray}
and the spin connection is evaluated as
\begin{eqnarray}
& &  \omega^{\hat{r}\hat{s}} \equiv \omega^{\hat{r}\hat{s}}_{\hat{\mu}}
dx^{\hat{\mu}} \;\Longrightarrow\; \omega^{r-} = \frac{1}{4}\pd^{r}G_{++} dx^{+}, \quad {\rm otherwise} = 0, \nn \\
& & \omega^{\hat{\mu}\hat{\nu}} \equiv e^{\hat{\mu}}_{\hat{r}}e^{\hat{\nu}}_{\hat{s}}\omega^{\hat{r}\hat{s}}\; \Longrightarrow\; \omega^{\mu -} = \frac{1}{4}\pd^{\mu}G_{++}dx^{+}, \quad {\rm otherwise} = 0. \nn 
\end{eqnarray}

\end{document}